\documentclass[9pt,twocolumn,twoside]{osajnl}

\journal{ol}

\setboolean{shortarticle}{true} 

\usepackage{graphicx}
\usepackage{multirow,bigdelim}
\usepackage{bm}
\usepackage{amsmath}
\usepackage{lipsum}

\usepackage[english]{babel}
\usepackage{blindtext}

\usepackage{color}

\newcommand{\mrm}{\mathrm}

\usepackage{booktabs}
\usepackage{array}
\newcolumntype{L}{>{\centering\arraybackslash}m{8cm}}

\title{Optimized ultra-narrow atomic bandpass filters via magneto-optic rotation in an unconstrained geometry
}
\date{\today}

\author[1,*]{James Keaveney}
\author[1]{Steven A. Wrathmall}
\author[1]{Charles S. Adams}
\author[1]{Ifan G. Hughes}

\affil[1]{Joint Quantum centre (JQC) Durham-Newcastle, Department of Physics, Durham University, South Road, Durham, DH1 3LE, United Kingdom}
\affil[*]{james.keaveney@durham.ac.uk}

\ociscodes{(020.7490) Zeeman effect; (120.2440) Filters; (020.1670) Coherent optical effects; (020.3690) Line shapes and shifts.}

\begin{abstract}
Atomic bandpass filters are widely used in a variety of applications, owing to their high peak transmission and narrow bandwidth. Much of the previous literature has used the Faraday effect to realize such filters, where an axial magnetic field is applied across the atomic medium. Here we show that by using a non-axial magnetic field, the performance of these filters can be improved in comparison to the Faraday geometry. We optimize the performance of these filters using a numerical model and verify their performance by direct quantitative comparison with experimental data. We find excellent agreement between experiment and theory. 
These optimized filters could find use in many of the areas where Faraday filters are currently used, with little modification to the optical setup, allowing for improved performance with relatively little change.
\end{abstract}

\setboolean{displaycopyright}{true}

\begin{document}

\maketitle



Magneto-optic effects in atomic media continue to be used in a vast array of applications, such as magnetometry~\cite{Kominis2003,Budker2007}, quantum hybrid systems integrating quantum dots with atomic media~\cite{Portalupi2016}, 
microwave detection and imaging~\cite{Horsley2015,Horsley2016,Alem2017}, optical isolators~\cite{Weller2012b} and self-stabilizing laser systems~\cite{Miao2011, Keaveney2016b, Chang2017}. 
%
%
Of particular interest are narrow optical bandpass filters~\cite{Ohman1956,Agnelli1975,Yeh1982,Dick1991,Menders1991,Zielinska2012,Zentile2015c,Zentile2015d}, the subject of this letter. Most of the current literature have used Faraday filters (often called FADOF filters), in which an axial magnetic field is applied along a medium realizing the Faraday effect~\cite{Faraday1846}. Faraday filters are widely used across many disciplines for atmospheric LIDAR~\cite{Fricke-Begemann2002}, velocimetry~\cite{Cacciani1978,Bloom1993}, optical communications~\cite{Junxiong1995}, quantum key distribution~\cite{Shan2006}, and laser frequency stabilisation~\cite{Miao2011,Keaveney2016b}, 
amongst others.

Atomic filters can be realized with ground- or excited-state transitions~\cite{Billmers1995,Zhang1998,Sun2011,Sun2012,Rudolf2012,Ling2014,Rudolf2014}, or can take advantage of pump-probe techniques in combination with the Faraday effect to create extremely narrow filters~\cite{Wang2012}, though at the cost of maximum transmission.
While most literature deals with weak signals, a recent investigation into strong-signal filters has also been reported~\cite{Luo2018}.
%
%
\begin{figure}[t!]
\centering
\includegraphics[width=0.9\columnwidth]{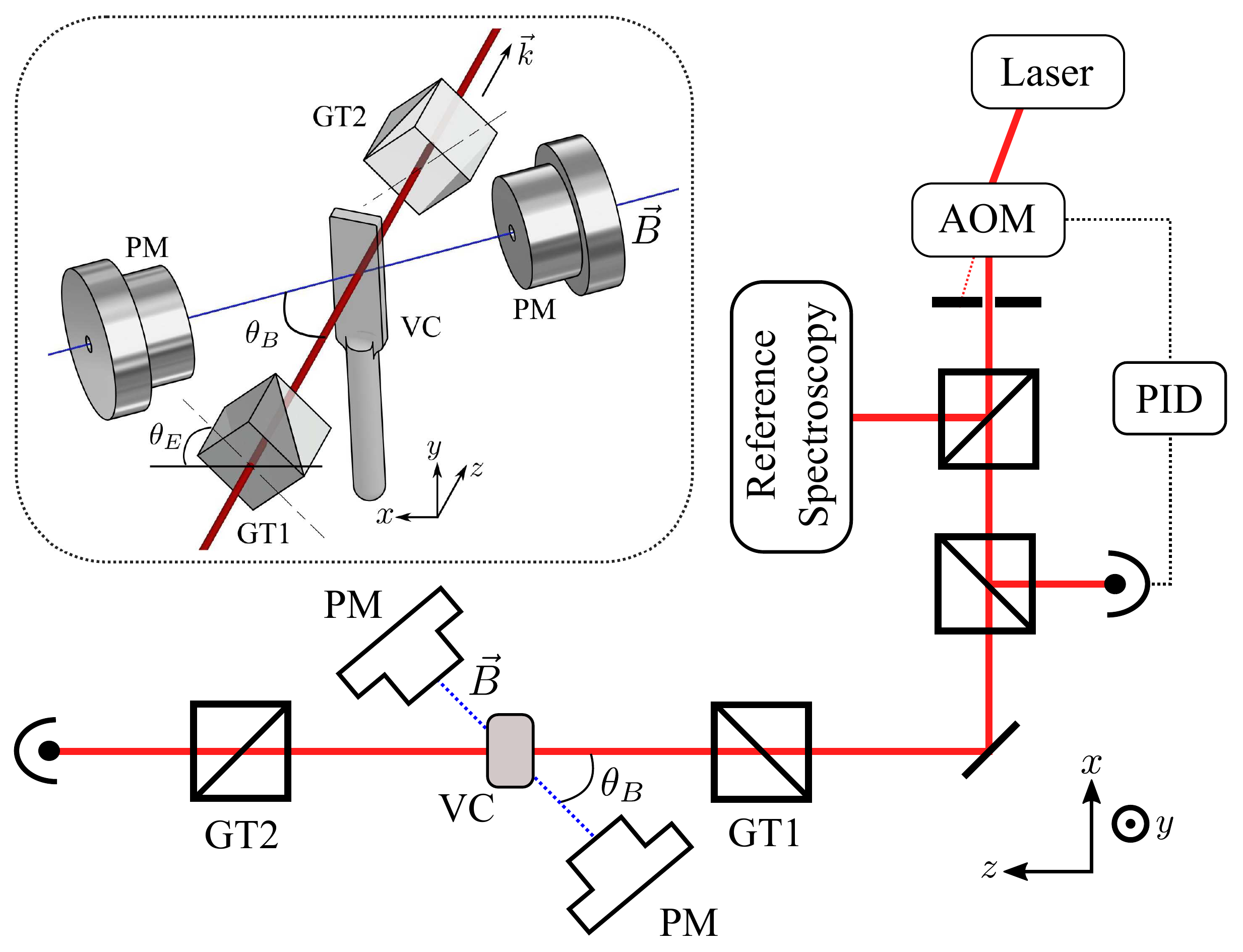}
\caption{Schematic of the magneto-optic filter setup and geometry. The filter is formed by placing an atomic vapor cell (VC) in an applied magnetic field ($B$) formed from two top-hat shaped permanent magnets (PM). The field strength is determined by the separation of the two magnets, and is adjustable up to 0.5~T. The magnetic field is oriented in the $xz$-plane at an angle $\theta_B$ to the $z$-axis, and sets the quantization axis for the atoms. The light propagates along the $z$-axis. An input high-extinction Glan-Taylor polarizer (GT1) is set at an angle $\theta_E$ with respect to the $x$-axis. The output polarizer is crossed at 90 degrees to the input polarizer.}
\label{fig:setup}
\end{figure}

The Faraday effect is conceptually and mathematically simple, 
but is only one special case of a more general magneto-optic phenomenon whereby an applied magnetic field causes birefringence and dichroism in an optically active medium. The two special cases, the Faraday effect and the Voigt effect~\cite{Voigt1898,Pershan1967,Schuller1991,Menders1992,Muroo1994}, occur when the magnetic field is aligned along or perpendicular to the light propagation axis, respectively. 
In systems with non-axial, non-perpendicular magnetic field geometries, no general elegant analytic solutions to the wave equation exist. It is therefore difficult to gain an intuitive understanding of light propagation through these media, and hence these general magneto-optic phenomena are discussed in the literature very rarely~\cite{Edwards1995,Nienhuis1998,Rotondaro2015}.

In this letter, we show that for many figures of merit, the performance of atomic optical filters is enhanced by allowing the magnetic field angle $\theta_B$ to be neither transverse nor axial. We numerically optimize relevant parameters for multiple atomic species using a recently developed computer model~\cite{Zentile2015b,Keaveney2017a} and validate these predictions with direct quantitative comparison to experimental data. We find excellent agreement between the theoretical model and experimental data, and demonstrate the best performance Rb D2 line bandpass filter measured to date.
%

%
%
\begin{figure}[t]
\begin{center}
\includegraphics[width=0.85\columnwidth,trim={0, 0.4cm, 0, 0.2cm}, clip=True]{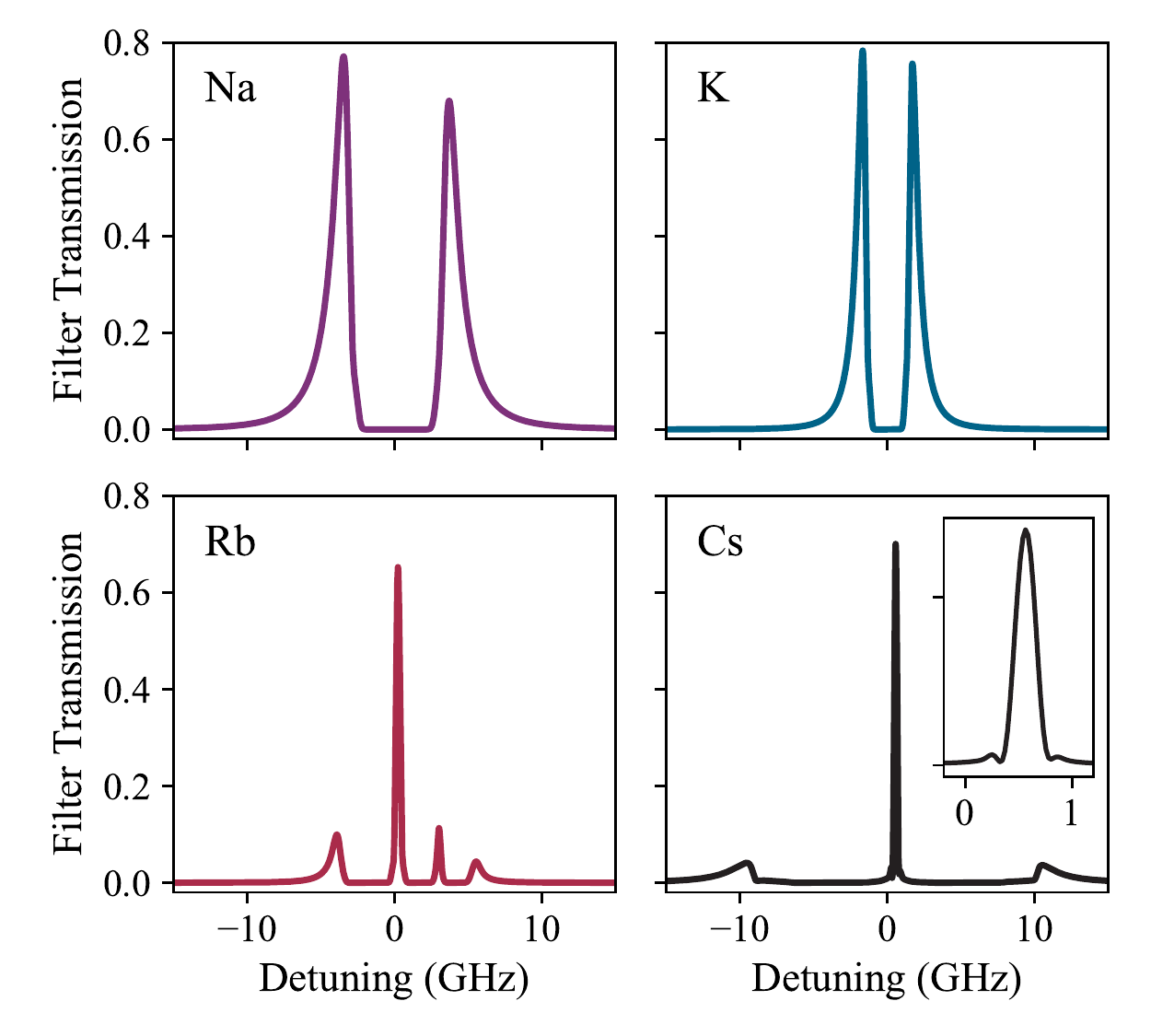}
\caption{Theoretical optimized filter profiles for the D2 lines of Na, K, Rb and Cs. The calculation parameters are displayed in table~1. The zero of the detuning axis represents the weighted line-center of the respective D2 lines~\cite{Siddons2008b,Zentile2015d}. Note that the K and Rb filters are for naturally abundant isotopic mixtures.
}
\label{fig:theoryplots}
\end{center}
\end{figure}
%

%
%
\begin{table*}[t]
\small
	\begin{center}
		\begin{tabular}{cccccccccc} 
\toprule
Element & D-line & T ($^\circ C$) & $\vert B \vert$ (G) & $\theta_B$ (deg.) & $\theta_E$ (deg.) & FWHM (GHz) & ENBW (GHz) & ${ \cal T}_{\rm max}$ (\%) & FOM (GHz$^{-1}$)
\\
\midrule
$^*$Rb & D2 & 126 & 230  & 83 & 6  & 0.34 & 0.63 & 65 & 1.04 \\
$^*$Rb & D1 & 147 & 695 & 90 & 45 & 0.62 & 1.37 & 78 & 0.57 \\
$^{85}$Rb & D2 & 121 & 224  & 83 & 90 & 0.34 & 0.72 & 67 & 0.93 \\
$^{85}$Rb & D1 & 139 & 310  & 90 & 46 & 0.43 & 0.98 & 62 & 0.63 \\
$^{87}$Rb & D2 & 121 & 849  & 83 & 81 & 0.36 & 0.99 & 69 & 0.69 \\
$^{87}$Rb & D1 & 141 & 315 & 90 & 45 & 0.40 & 0.76 & 61 & 0.80 \\
Cs & D2 & 127 & 1120  & 87 & 89 & 0.24 & 0.56 & 70 & 1.24 \\
Cs & D1 & 121 & 338 & 89 & 46 & 0.36 & 0.70 & 68 & 0.97 \\
$^*$K & D2 & 150 & 88  & 1 & 87 & 0.72 & 1.90 & 78 & 0.41 \\
$^*$K & D1 & 177 & 460  & 90 & 47 & 0.65 & 1.63 & 76 & 0.47 \\
Na & D2 & 245 & 144  & 0 & 0 & 1.33 & 3.34 & 77 & 0.23 \\
Na & D1 & 279 & 945 & 88 &  41 & 1.30 & 3.32 & 79 & 0.24 \\
\bottomrule
\vspace{-0.5cm}
\label{t:optresults}
\end{tabular}
\caption{optimized filter parameters for 5 mm vapor cell thickness across the commonly used alkali-metal atoms. $*$ = using natural isotopic abundance ratios.}
\end{center}
\end{table*}

Whilst evaluating the performance of optical filters is application dependent, there are some general figures of merit which are widely used. The equivalent noise bandwidth (ENBW) is one such measure, defined as
\begin{equation}
\mrm{ENBW}=\frac{\int^\infty_0 \mathcal{T}(\omega)\mrm{d}\omega}{\mathcal{T}(\omega_s)},
\label{eq:ENBW}
\end{equation} 
where $\cal{T}$ is the light intensity after the filter, $\omega$ is the angular optical frequency and $\omega_s$ is the signal angular frequency. If a specific signal frequency is not required, the signal frequency can be set to the frequency which gives the maximum transmission ($\mathcal{T}(\omega_s)=\mathcal{T}_{\rm max}$).

One is usually interested in reducing the ENBW, but if this is done with computer minimization routines the returned solution is a filter which has zero transmission at all frequencies. Many applications require the combination of high peak transmission with a narrow bandwidth, so a better figure-of-merit (FOM) is
\begin{equation}
\mrm{FOM} = \left.\frac{\mathcal{T}_\mrm{max}^2}{\int^\infty_0 \mathcal{T}(\omega)\mrm{d}\omega} = \frac{\mathcal{T}_\mrm{max}}{\mrm{ENBW}}\right\rvert_{\mathcal{T}{(\omega_s)}=\mathcal{T}_\mrm{max}},
\label{eq:FOM1}
\end{equation}
as first suggested in ref.~\cite{Kiefer2014}. Using this FOM, we can maintain reasonably large transmission while minimizing ENBW. 
This FOM has previously been used to find the optimal performance of atomic Faraday filters~\cite{Zentile2015c,Zentile2015d}, and in the case of an unconstrained magnetic field angle to theoretically predict improved performance on the Cs D2 line~\cite{Rotondaro2015}.

Fig.~\ref{fig:setup} shows a schematic of the optical setup, and the geometry of the situation. 
A laser beam propagates through an atomic vapor cell (VC) in the presence of a magnetic field. The magnetic field vector at the position of the vapor cell is oriented in the $xz$-plane at an angle $\theta_B$ to the $z$-axis. Values of $\theta_B=0$ and $\theta_B=90^\circ$ yield the Faraday and Voigt geometries, respectively.
The cell is placed between two high-extinction polarizers. The coupling between atomic transitions and polarization modes of the light is dependent on the angle between the electric field vector of the light and the magnetic field vector of the applied field. 
Therefore, the angle of the input polarizer, $\theta_E$, is variable, but the relative angle of the two polarizers GT1 and GT2 is always 90 degrees (i.e. crossed-polarizers), such that in the absence of any atom-light interaction, the transmission is maximally extinguished.

In the Faraday geometry, the Faraday effect arises from a difference in the refractive indices that couple to the circular polarization components. This leads to a relative phase change and hence a polarization rotation on propagation through such a medium.
When the angle between the light propagation axis, $\vec{k}$, and the magnetic field vector $\vec{B}$ are unconstrained, one must solve the wave equation to find the normal modes of propagation in the system~\cite{Palik1970,Rotondaro2015,Keaveney2017a}. Like in the Faraday case, there are two normal modes associated with two refractive indices, but these are not trivial to calculate and in general the solutions must be computed numerically. The coupling to polarization components of light is again non-trivial, and while there is still a magneto-optic rotation it does not have an intuitive description.
We have developed a computer model, ElecSus~\cite{Zentile2015b,Keaveney2017a}, which solves this problem and allows for calculation of the transmission of a weak-probe beam~\cite{Sherlock2009} through an alkali-metal atomic vapor with a magnetic field of any strength and angle, and with an arbitrary transverse input electric field polarization. We apply this model to Na, K, Rb and Cs atoms, on the D1 and D2 lines ($nS_{1/2} \rightarrow nP_{1/2}, nP_{3/2}$ transitions, respectively) and calculate a filter profile as a function of the laser frequency detuning from resonance based on the arrangement of optics shown in fig.~\ref{fig:setup}.

Computer optimization in the Faraday geometry has been previously demonstrated~\cite{Kiefer2014,Zielinska2014,Zentile2015c,Zentile2015d}, but in the unconstrained geometry the problem is more computationally difficult. In addition to some fixed parameters (element and relative isotopic abundance, cell length, and D-line), there are four parameters the optimization routine can vary. These are the cell temperature $T$, which sets the atomic number density and hence the amount of atom-light interaction; the magnetic field strength $\vert B \vert$, magnetic field angle $\theta_B$ and initial polarization angle $\theta_E$. 
This forms a complex multi-dimensional parameter space, with many local minima - finding the global minimum of such a system is highly non-trivial.
Our method takes a randomized set of trial parameters for which we evaluate the FOM. From these trial parameters we take a small fraction (the ones with the highest FOM) and run a downhill optimization algorithm using the trial parameters as initial parameters. After each trial has optimized we take the solution with the best FOM as the best estimate of the global minimum.

The choice of vapor cell length can drastically affect the optimized performance of the optical filter, as has been previously demonstrated~\cite{Zentile2015c}. Experimentally, one needs to balance the optical depth requirements with the ability to produce an approximately uniform magnetic field across the cell. It is easier to create a uniform magnetic field over a smaller volume, but reducing the cell thickness requires a higher vapor density to compensate for the loss of optical depth, causing additional broadening which, in the weak-probe regime, negatively affects filter performance~\cite{Zentile2015d} (although additional broadening in high power filters can be beneficial, as demonstrated recently~\cite{Xiong2018}). In this work we choose a cell thickness of 5~mm, which represents a compromise between these two factors, and also matches a vapor cell length that is available to us in the experiment.

Table~1 
shows results for the commonly used alkali-metal atoms Na, K, Rb and Cs, and isotopically pure $^{85}$Rb and $^{87}$Rb which are commonly available vapor cells. In all cases (apart from the Na D2 line, which optimizes to the Faraday geometry), the FOM is better than equivalent filters for the Faraday geometry (see table 1 of ref.~\cite{Zentile2015c} for comparison, though note this previous work optimized for a 75 mm cell). In particular, the Rb D2 line is better than the Faraday case by close to a factor of 3 in FOM.
In fig.~\ref{fig:theoryplots} we show examples of optimized profiles for the D2 lines of Na, K, Rb (naturally abundant) and Cs. For Rb and Cs we find an optimum filter operating in the `line-center' mode~\cite{Yeh1982,Zentile2015d}, whereas for Na and K we find a `wing-type' filter~\cite{Yeh1982,Zentile2015d}.
Of all the results, the Cs D2 line filter has the highest FOM value of 1.24~GHz$^{-1}$. However, as previously discussed, the cell length makes a large difference to the filter properties. If we increase the cell thickness to 75~mm, we calculate an optimized FOM value of 2.43~GHz$^{-1}$, with a broadly similar spectral profile to that shown in figure~\ref{fig:theoryplots} but with a FWHM of just 110~MHz. This solution has a much larger FOM than the equivalent filter predicted in ref.~\cite{Rotondaro2015}. The difference in result could be due to our computational optimization approach which has a higher success of finding the global solution, rather than falling into a local optimum.


We demonstrate the validity of our model via a direct quantitative comparison with experimental data.
In the experiment, two top-hat shaped permanent (NdFeB) magnets create a magnetic field whose strength is adjustable via the magnets' separation up to 0.5~T. The local magnetic field in the vicinity of the vapor cell is approximately uniform; the change over the interrogated volume is less than 1\% in strength and 1 degree in direction.
As in the model, the field is oriented in the $xz$-plane at an angle $\theta_B$ with respect to the light propagation $z$-axis. The magnets are mounted on a large rotation platform and can be oriented at nearly any angle. The applied magnetic field defines the quantization axis for the atoms.
The vapor cell has an optical path length of 5~mm, is filled with Rb in its natural abundance ratio, and is heated to provide sufficient vapor pressure and hence atomic number density. A weak-probe~\cite{Sherlock2009} beam (100 nW optical power, focused in the vapor cell to a 1/e$^2$ waist of approximately 100~$\mu$m) is sent through the optical filter and detected on a photodiode as the laser is scanned across the Rb D2 resonance lines. Reference spectroscopy comprising a Fabry-Perot etalon and a 75~mm room temperature zero-field atomic reference cell are used to calibrate the frequency axis of the laser, following ref.~\cite{Keaveney2014a}, while the optical power is actively controlled via a power monitor photodiode and a feedback loop to the RF power of an acousto-optic modulator, as in ref.~\cite{Truong2012}. This results in a stable optical power level as the laser is scanned across the Rb D2 resonance lines - the filter transmission is therefore easily normalized by taking a reference transmission level, obtained by rotating the output polarizer GT2 so that the transmission is maximized.

\begin{figure}[t]
\begin{center}
\includegraphics[width=0.9\columnwidth,trim={0, 0.4cm, 0, 0.4cm}, clip=True]{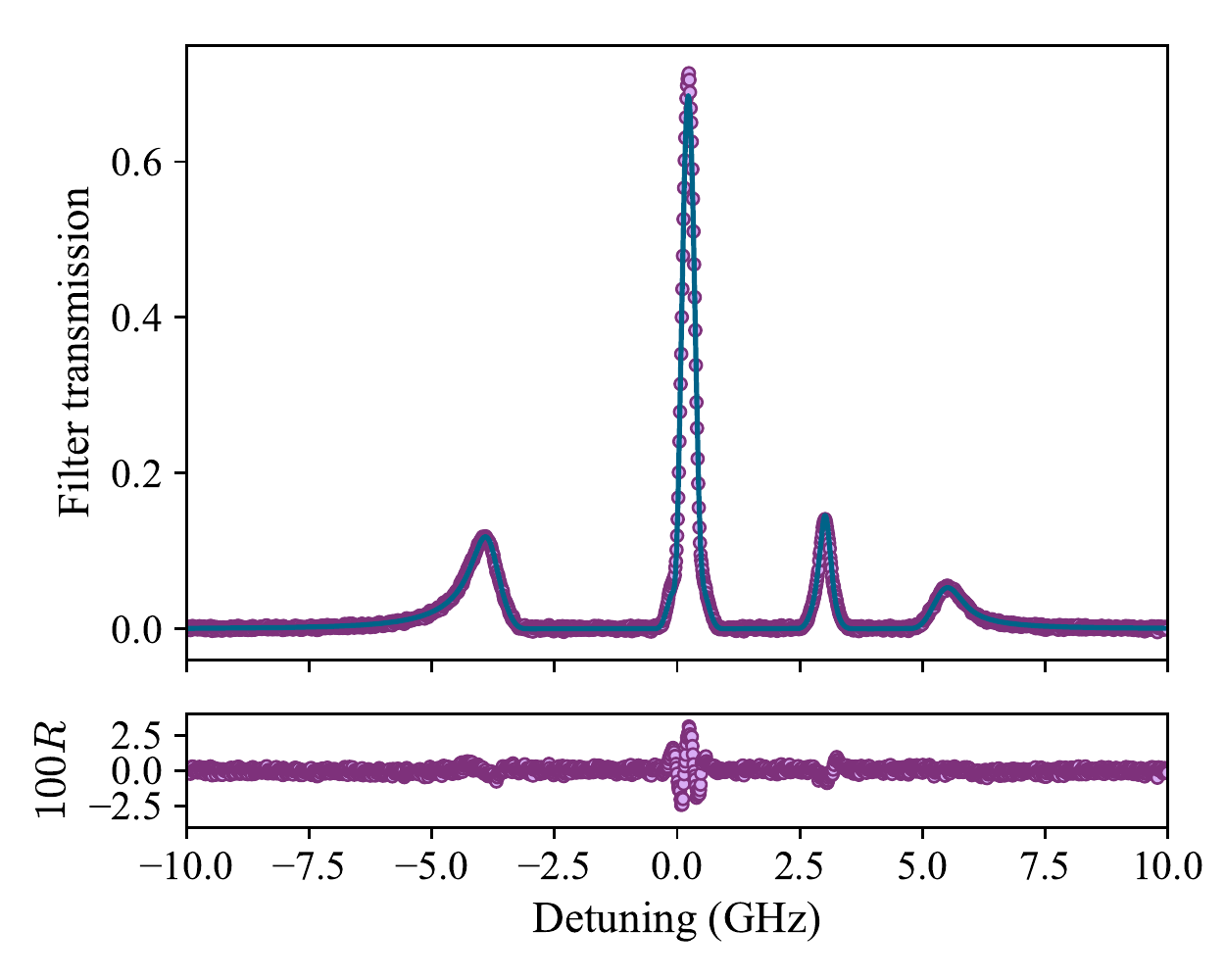}
\caption{Comparison of experimental data with theoretical model for the Rb D2 line (natural abundance ratio). The experimental optical path length is 5~mm. Purple points are experimental data, and the blue solid line is the fit to the model with fitted parameters $T = 124.0^\circ$C, $\vert B \vert = 232$~G, $\theta_B = 81.8^\circ$, $\theta_E = 2.9^\circ$. The experimentally determined ENBW and FOM are $(0.68\pm0.01)$~GHz and $(1.04\pm0.01)$~GHz$^{-1}$, respectively.}
\label{fig:expt_vs_theory}
\end{center}
\end{figure}

Fig.~\ref{fig:expt_vs_theory} shows the results of the experiment (purple data points). We fit the data with the same model used to predict the optimal parameters; the magnetic field strength and angle, temperature of the cell and input polarization angle are floating parameters, allowing for small imperfections in the experimental setup. We find excellent agreement between theory and experiment, with a RMS error between experiment and theory of 0.35\% and almost structureless residuals~\cite{Hughes2010}. The fit parameters $T = 124.0^\circ$C, $\vert B \vert = 232$~G, $\theta_B = 81.8^\circ$, $\theta_E = 2.9^\circ$ are within 1\% (B, T) or 2 degrees (angles) of the experimentally measured parameters, and are very close to the theoretical optimum parameters. 
The experimental ENBW of $(0.68\pm0.01)$~GHz and FOM of $(1.04\pm0.01)$~GHz$^{-1}$ represent the narrowest high transmission optical filter demonstrated in a Rb vapor to date.
%
%
%

In conclusion, we have developed a theoretical model to predict the optimal performance of optical filters in atomic vapors. We find that by relaxing the constraint of the magnetic field angle, with respect to the light propagation axis, the performance as measured by a figure-of-merit can be greatly increased over what can be achieved in the Faraday geometry. This enhancement of performance has been experimentally verified, and we find excellent agreement between the model and experimental data. In addition, we predict an improved Cs filter in a 75~mm vapor cell with an even higher FOM of 2.43~GHz$^{-1}$ and narrower bandwidth than previously calculated. These unconstrained geometry filters can find use in many areas where Faraday filters are already used, with only slight changes to experimental configurations. We envisage further performance gains may be possible by combining these atomic bandpass filters with other atomic media which would act as notch filters to remove some of the side-band peaks whilst retaining the high peak transmission, or cascading multiple vapor cells under different conditions such that the optical rotation combines favorably for enhanced filter performance. These avenues will form the basis of future investigations.


The authors acknowledge funding from EPSRC (Grant Nos. EP/L023024/1 and EP/R002061/1). This work made use of the facilities of the Hamilton HPC Service of Durham University.
The data presented in this paper are available from DRO\footnote{\url{DOI://xxx.xxxxxxx}, added at proof stage}.
 
\bibliography{library,library2}

\end{document}